\documentclass[10pt,reqno,twoside,paper=letter]{amsart}

\usepackage[letterpaper,left=20mm,right=14mm,layoutvoffset=10mm]{geometry}
\usepackage{epsfig,graphicx}

\usepackage[bf,small]{titlesec}
\titlelabel{\thetitle.\,\,\,}
\usepackage{amssymb,amsfonts}
\usepackage{amsmath}
\usepackage{float} 
\usepackage{multicol}
\usepackage{fancyhdr}
\usepackage[latin1]{inputenc}				
\usepackage[comma,longnamesfirst,sectionbib]{natbib}
\usepackage{enumerate}
\usepackage[symbol*]{footmisc}
\DeclareMathOperator{\acot}{acot}
\newcommand{\inforevista}{\scriptsize  Rev. Acad. Colomb. Cienc. Ex. Fis. Nat. nn(nnn):ww--zzz,ddd-ddd de 2016}
\begin{document}
%
%
\pagenumbering{arabic}
\fancypagestyle{plain}{%
\fancyhf{} 
\fancyfoot[R]{\thepage} %
\fancyhead[L]{\inforevista}
\renewcommand{\headrulewidth}{0pt}
\renewcommand{\footrulewidth}{0pt}}
\thispagestyle{plain} 
\pagestyle{fancy}     
\fancyhead{} 
\renewcommand{\headrulewidth}{0.0pt} 
\fancyhead[LO]{\inforevista}
\fancyhead[RO]{\scriptsize Analytical Potentials for Flat Galaxies with Spheroidal Halos}
\fancyhead[LE]{\scriptsize G. A. González and J. I. Reina}
\fancyhead[RE]{\inforevista}
\fancyfoot{} 
\fancyfoot[LE]{\thepage}
\fancyfoot[RO]{\thepage}

\begin{flushright}
\rule[-0.5ex]{0.5ex}{3.0ex} {\large Physical Sciences}
\end{flushright}
\vspace*{0.3cm}

\begin{center}
{\LARGE \textbf{Analytical Potentials for Flat Galaxies with Spheroidal Halos \\}}
\end{center}

\vspace{3mm}
\begin{center}
\textbf{\small Guillermo A. González${}^{1,}\footnotemark[1], $ and Jerson I. Reina${}^{1,2}$}
\vspace{3mm}

{\scriptsize
${}^{1}$Escuela de Física, Universidad Industrial
de Santander, Bucaramanga, Colombia\\
${}^{2}$Departamento de Ciencias B\'asicas, Universidad Santo Tom\'as-Bucaramanga, Bucaramanga, Colombia\\
 }
\end{center}
\footnotetext[1]{Correspondence: G. A. González, guillermo.gonzalez@saber.uis.edu.co, Received xxxxx XXXX; Accepted xxxxx XXXX.}

\begin{Small}
\vspace{3.0mm}
\rule{\textwidth}{0.4pt}

\begin{center}
\begin{minipage}{14cm}
\vspace{3mm}

\textbf{Abstract}
\vspace{3mm}

{A family of analytical potential-density pairs for flat galaxies with spheroidal halos is presented. The potential are obtained by means of the sum of two independent terms: a potential associated with a thin disc and a potential associated with a spheroidal halo, which are expressed as appropriated superpositions of products of Legendre functions, in such a way that the model implies a linear relationship between the masses of the thin disc and the spheroidal halo. By taking a particular case for the halo potential, we found that the circular velocity obtained can be adjusted very accurately to the observed rotation curves of some specific galaxies, so that the models are stable against radial and vertical perturbations. Two particular models for the galaxies NGC4389 and UGC6969 are obtained by adjusting the circular velocity with data of the observed rotation curve of some galaxies of the Ursa Mayor Cluster, as reported in \citet{VS}. The values of the halo mass and the disc mass for these two galaxies are computed obtaining a very narrow interval of values for these quantities. Furthermore, the values of obtained masses are in perfect agreement with the expected order of magnitude and with the relative order of magnitude between the halo mass and the disc mass.
\\[1mm]

}

\vspace{3mm}

\textbf{Potenciales Anal\'iticos Para Galaxias Planas con Halos Esferoidales}
\vspace{3mm}

\textbf{Resumen}
\vspace{3mm}

{Se presenta una familia de pares  anal\'iticos potencial-densidad para galaxias planas con halos esferoidales. Los potenciales son obtenidos por medio de la suma de dos t\'erminos independientes: un potencial asociado al disco delgado y un potencial asociado al halo esferoidal, los cuales son expresados apropiadamente como la superposici\'on de productos de funciones de Legendre, de tal manera que el modelo implica una relaci\'on lineal entre las  las masas del disco delgado y el halo esferoidal. Tomando un caso particular para el potencial del halo, encontramos que la velocidad circular obtenida puede ser ajustada muy precisamente con la curva de rotaci\'on de algunas galaxias espec\'ificas, de tal manera que los modelos son estables contra perturbaciones radiales y verticales. Dos modelos particulares para las galaxias NGC4389 y UGC6969 son obtenidos ajustando la velocidad circular del modelo con datos de la curva de rotación observada de algunas galaxias del Cluster de la Osa Mayor, reportados en \cite{VS}. Los valores de la  masa del halo y la masa del disco para estas dos galaxias son calculados obteniendo un intervalo muy estrecho de valores para dichas cantidades. Adem\'as, los valores de masa aqu\'i obtenidos est\'an en perfecto acuerdo con el orden de magnitud esperado y con el orden de magnitud relativo entre la masa del halo y la masa del disco.
\\[1mm]

\textbf{Palabras clave:} Teor\'ia del Potencial, Galaxias de Disco, Mec\'anica Celeste, Masa de Galaxias.}
\end{minipage}
\end{center}

\rule{\textwidth}{0.4pt}
\end{Small}

\begin{small}
\columnsep 0.5 cm
\begin{multicols}{2}

\setlength{\parskip}{.3cm}

\section*{Introduction}

One of the oldest and most important problems in galactic dynamics is the 
determination of the mass distribution based on the observations of the circular 
velocity or rotation curve \citep{PIER}, defined as the speed of the stars moving 
in the galactic plane in circular orbits around the center. Now, if we assume a 
particular model for the composition of the galaxy, the fit of that model with 
the rotation curve of a particular galaxy can, in principle, completely determine 
the distribution of mass. So then, the rotation curve provides the most direct 
method to measure the distribution of mass of a galaxy \citep{BT}.

Currently, the most accepted description of the composition of spiral galaxies is 
that a significant portion of its mass is concentrated in a thin disc, while the 
other contributions to the total mass of the galaxy come from a spherical halo 
of dark matter, a central bulge and, perhaps, a central black hole \citep{BT}. 
Now, since all components contribute to the gravitational field of the galaxy, 
obtaining appropriate models that include the effects of all parts is a problem 
of great difficulty. However, the contribution of each part is limited to 
certain distance scales, so in a reasonably realistic model it is not necessary to 
include the contribution of all components \citep{FAB}.

In particular, the gravitational influence of the central black hole is 
appreciable only within a few parsecs around the center of the galaxy 
\citep{NAT}, so it can be completely neglected when studying the dynamics of the 
disc, or in regions outside the central bulge, while the bulge mainly dominates 
the inner region of the galaxy to a few kiloparsec. So then, the main 
contributions to the gravitational field of the galaxy come from the galactic 
disc and the dark matter halo \citep{FAB}. However, it is commonly accepted that 
many aspects of galactic dynamics can be described, in a fairly approximate way, 
using models that consider only the contribution of a thin galactic disc 
\citep{BT}.

Accordingly, the study of the gravitational potential generated by an idealized
thin disc is a problem of great astrophysical relevance and so, through the
years, different approaches have been used to obtain such kind of thin disc
models (see \citet{BT} and references therein). So, once an expression for the 
gravitational potential has been derived, corresponding expressions for the 
surface mass density of the disc and for the circular velocity of the disc 
particles can be obtained. Then, if the expression for the circular velocity can 
be adjusted to fit the observational data of the rotation curve of a particular 
galaxy, the total mass can be obtained by integrating the corresponding surface 
mass density.

However, although most of these thin disc models have surface densities and
rotation curves with remarkable properties, many of them mainly represent discs
of infinite extension and thus they are rather poor flat galaxy models.
Therefore, in order to obtain more realistic  models of flat galaxies, it is better
to consider methods that permit obtaining finite thin disc models. Now, a simple
method to obtain the gravitational potential, the surface density and the
rotation curve of thin discs of finite radius was developed by \citet{HUN}, the
simplest example of a disc obtained by this method being the well known
\citet{KAL} disc.

In a previous paper \citep{GR} we used the Hunter method in order to obtain an 
infinite family of thin discs of finite radius with a well-behaved surface mass 
density. This family of disc models was derived by requiring that the surface 
density behaves as a monotonously decreasing function of the radius, with a 
maximum at the center of the disc and vanishing at the edge. Furthermore, the 
motion of test particles in the gravitational fields generated by the first four 
members of this family was studied in \citet{RLG}. So, although the mass 
distribution of this family of discs presents a satisfactory behaviour in
such a way that they could be considered adequate as flat galaxy models, their
corresponding rotation curves do not present a so good behavior, as they do not
reproduce the flat region of the observed rotation curve.

On the other hand, in \citet{PRG} a new family of discs was obtained as a
superposition of members of the previously obtained family, by requiring that the
surface density be expressed as a well-behaved function of the gravitational
potential, in such a way that the corresponding distribution functions can be
easily obtained. Furthermore, besides presenting a well-behaved surface density,
the models also presented rotation curves with a better behavior than the
generalized Kalnajs discs. However, although these discs are stable against
small radial perturbations of disc star orbits, they are unstable to small
vertical perturbations normal to the disc plane. Then, apart from the stability problems,
these discs can be considered as quite adequate models in order to satisfactorily describe a great variety of galaxies.

Based on these works, in \citet{GPR} were obtained some thin disc models in which 
the circular velocities were adjusted to very accurately fit the observed 
rotation curves of four spiral galaxies of the Ursa Major cluster, galaxies 
NGC3877, NGC3917, NGC3949 and NGC4010. These models presented well-behaved 
surface densities and the obtained values for the corresponding total mass agree with the expected order of magnitude. However, the models presented a 
central region with strong instability to small vertical perturbations. Now,
this result was expected as a consequence of the fact that the models only 
consider the thin galactic disc. Therefore, more realistic models must be 
considered including the non-thin character of the galactic disc or the
mass contribution of the spheroidal halo.

In agreement with the above considerations, in this paper we will consider a 
family of models obtained by expressing the gravitational potential as the 
superposition of a potential generated by the thin galactic disc and a potential 
generated by the spheroidal halo, in such a way that the model implies 
a linear relationship between the masses of the thin disc and the spheroidal 
halo. By adjusting the corresponding expression for the circular velocity to the 
observed data of the rotation curve of some specific galaxies, some particular 
models will be analysed. Then, from the corresponding expressions for the disc 
surface density and the density of the halo, estimate values for the total mass 
of the disc and the total mass of the halo will be obtained. The paper is organised as follows. First we present the thin disc plus halo model. Then, we obtain the corresponding expressions for particular models, and then the models are fitted to data of the observed rotation curve of some galaxies of the Ursa Mayor Cluster, as reported in \citet{VS}. Finally, we discuss the obtained results.

\section*{The Thin Disc Plus Halo Model}\label{dischalo}

In order to obtain galaxy models consisting of a thin galactic disc and a
spheroidal halo, we begin considering an axially symmetric gravitational
potential $\Phi = \Phi(R,z)$, where $(R,\varphi,z)$ are the usual cylindrical
coordinates. Also, besides the axial symmetry, we suppose that the potential 
has symmetry of reflection with respect to the plane $z = 0$,
\begin{equation}
\Phi (R,z) = \Phi (R,-z), \label{parity}
\end{equation}
which implies that the normal derivative of the potential satisfies the relation
\begin{equation}
\frac{\partial \Phi}{\partial z} (R,-z) = - \frac{\partial \Phi}{\partial z}
(R,z), \label{reflect}
\end{equation}
in agreement with the attractive character of the gravitational field.  We also
assume that $\partial \Phi / \partial z$ does not vanish on the plane $z = 0$,
in order to have a thin distribution of matter that represents the disc.

On the other hand, in order to separately describe the thin disc and the
spheroidal halo, we consider that the gravitational potential can be written as
the superposition of two independent components
\begin{equation}
\Phi (R,z) = \Phi_d (R,z) + \Phi_h (R,z), \label{poten}
\end{equation}
where $\Phi_d (R,z)$ is the part of the potential generated by the thin galactic
disc, while $\Phi_h (R,z)$ corresponds to the spheroidal halo component. The 
disc component $\Phi_d (R,z)$ must be a solution of the Laplace equation 
everywhere outside the disc,
\begin{equation}
\nabla^2 \Phi_d = 0, \label{lapec}
\end{equation}
while the halo component $\Phi_h (R,z)$ satisfies the Poisson equation
\begin{equation}
\nabla^2 \Phi_h = 4 \pi G \varrho, \label{posec}
\end{equation}
where $\varrho (R,z)$ is the mass density of the halo.

So, given a potential $\Phi(R,z)$ with the previous properties, we can easily
obtain the circular velocity $v_c (R)$, defined as the velocity of the stars
moving at the galactic disc in circular orbits around the center, through the
relationship
\begin{equation}
v_{c}^{2}(R) =  R \left. \frac{\partial \Phi}{\partial R} \right|_{z=0},
\label{vc2} 
\end{equation}
while the surface mass density $\Sigma(R)$ of the thin galactic disc is given by
\begin{equation}
\Sigma(R) = \left. \frac{1}{2 \pi G} \frac{\partial \Phi}{\partial z}
\right|_{z=0^{+}},
\end{equation}
which it is obtained by using the Gauss law and the reflection symmetry of
$\Phi(R,z)$.

Accordingly, in order that the potential of the spheroidal halo does not
contribute to the disc surface density, we will impose the condition
\begin{equation}
\left. \frac{\partial \Phi_h}{\partial z} \right|_{z=0^{+}} = 0.
\label{cond1}
\end{equation}
Furthermore, in order to have a surface density corresponding to a finite
disclike distribution of matter, we impose boundary conditions in the form
\begin{subequations}
\begin{align}
\left. \frac{\partial \Phi_d}{\partial z} \right|_{z=0^{+}} \neq 0;
\quad R \leq a, \label{cond2}\\
\left. \frac{\partial \Phi_d}{\partial z} \right|_{z=0^{+}} = 0; \quad R > a,
\label{cond3}
\end{align}
\end{subequations}
in such a way that the matter distribution is restricted to the disc $z = 0$, $0
\leq R \leq a$, where $a$ is the radius of the disc.

In order to properly pose the boundary value problem, we introduce the oblate
spheroidal coordinates, whose symmetry adapts in a natural way to the geometry
of the model. These coordinates are related to the usual cylindrical coordinates
by the relation \citep{MF}
\begin{subequations}
\begin{align}
R &= a \sqrt{(1+\xi^{2})(1-\eta^{2})},\\
z &= a \xi \eta,
\end{align}
\end{subequations}
where $0 \leq \xi < \infty$ and $-1 \leq \eta < 1$.  The disc has the
coordinates $\xi=0$, $0 \leq \eta^{2} < 1$.  On crossing the disc, the $\eta$
coordinate changes sign but does not change in absolute value.  The singular
behaviour of this coordinate implies that an even function of $\eta$ is a
continuous function everywhere but has a discontinuous $\eta$ derivative at the
disc.

Now, in terms of the oblate spheroidal coordinates, the Laplace operator acting
over any axially symmetric function $\Phi ( \xi,\eta)$ gives
\begin{equation}
\nabla^2 \Phi = \frac{\left[(1 + \xi^{2}) \Phi_{,\xi} \right]_{,\xi} +
\left[(1 - \eta^{2}) \Phi_{,\eta} \right]_{,\eta}}{a^2 (\xi^2 + \eta^2)},
\label{lapop}
\end{equation}
whereas the boundary condition (\ref{cond1}) is equivalent to
\begin{subequations}
\begin{align}
\left. \frac{\partial \Phi_h}{\partial \xi} \right|_{\xi = 0} &= 0,
\label{cond1a} \\
\left. \frac{\partial \Phi_h}{\partial \eta} \right|_{\eta = 0} &= 0,
\label{cond1b}
\end{align}
\end{subequations}
and the boundary conditions (\ref{cond2}) and (\ref{cond3}) reduce to
\begin{subequations}
\begin{align}
\left. \frac{\partial \Phi_d}{\partial \xi} \right|_{\xi = 0} &\neq 0,
\label{cond2a} \\
\left. \frac{\partial \Phi_d}{\partial \eta} \right|_{\eta = 0} &= 0.
\label{cond2b}
\end{align}
\end{subequations}
Moreover, in order for the gravitational potential to be continuous everywhere,
$\Phi (\xi,\eta)$ must be an even function of $\eta$, which grants also the
fulfilment of conditions (\ref{cond1b}) and (\ref{cond2b}).

Accordingly, by imposing the previous boundary conditions over the general
solution of the Laplace equation in oblate spheroidal coordinates, we can write
the gravitational potential of the galactic disc as \citep{BAT}
\begin{equation}
\Phi_n (\xi,\eta) = - \sum_{l=0}^{n} C_{2l}\ q_{2l}(\xi) P_{2l}(\eta),
\label{potend}
\end{equation}
where $n$ is a positive integer, which it defines the model of disc 
considered. Here $P_{2l}(\eta)$ are the usual Legendre polynomials and $q_{2l}
(\xi)=i^{2l+1}Q_{2l}(i\xi)$, with $Q_{2l}(x)$ the Legendre functions of second 
kind (see \cite{AW} and, for the Legendre functions of 
imaginary argument, \cite{MF}, page 1328). The coefficients 
$C_{2l}$ are, in principle, arbitrary constants, though they must be specified to 
obtain any particular model. We will do this later on, by adjusting the circular 
velocity of the model with the observed data of the rotation curve of some 
specific galaxies.

With this expression for the gravitational potential of the disc, the surface
density is given by
\begin{equation}
\Sigma({\widetilde R}) = \frac{1}{2\pi a G \eta}\sum_{l=0}^{n} C_{2l} (2l+1)
q_{2l+1}(0) P_{2l}(\eta), \label{density}
\end{equation}
where, as $\xi = 0$, $\eta = \sqrt{1 - {\widetilde R}^2}$, with ${\widetilde R}
= R/a$. Then, by integrating on the total area of the disc, we find the value
\begin{equation}
\frac{\mathcal{M}_d G}{a} = C_0\label{masa}
\end{equation}
for the total mass of the disc. Now, it is clear that the surface density 
diverges at the disc edge, when $\eta = 0$, unless that we impose the condition 
\citep{HUN}
\begin{equation}
\sum_{l = 0}^{n} C_{2l} (2l+1) q_{2l+1}(0) P_{2l}(0) = 0, \label{finitden}
\end{equation}
that, after using the identities
\begin{subequations}
\begin{align}
&P_{2n} (0) = (-1)^n \frac{(2n - 1)!!}{(2n)!!}, \\
&q_{2n+1} (0) = \frac{(2n)!!}{(2n + 1)!!},
\end{align}
\end{subequations}
which are easily obtained from the properties of the Legendre functions, leads
to the expression
\begin{equation}
C_0 = \sum_{l = 1}^{n} (-1)^{l + 1} C_{2l}, \label{constant0}
\end{equation}
which gives, through (\ref{masa}), the value of the disc mass
$\mathcal{M}_d$ in terms of the constants $C_{2l}$, with $l \geq 1$.

Now, to properly choose the gravitational potential of the spheroidal halo, we 
consider the superposition
\begin{equation}
\Phi_h (\xi,\eta) = \sum_{j=0}^{m} \sum_{k=0}^{j} B_{jk} \ q_{j}^{k} (\xi) P_{j}^{k} (\eta), \label{pothgen}
\end{equation}
where $m$ is a positive integer, which defines the model of halo considered, and the 
coefficients  $B_{jk}$ are arbitrary constants which must be specified to obtain any 
particular model. Here \citep{LAM}
\begin{equation}
q_{j}^{k} (\xi) = (1 + \xi^2)^{\frac{k}{2}} \frac{d^k q_{j}(\xi)}{d\xi^k},
\label{qjk}
\end{equation}
are the solutions of the differential equation
\begin{equation}
\frac{d }{d\xi} \left[(1 + \xi^2) \frac{dq_{j}^{k}}{d\xi} \right] = \left[j(j+1)
- \frac{k^2}{1 + \xi^2}\right] q_{j}^{k} (\xi), \label{dqjk}
\end{equation}
while the associated Legendre functions \citep{AW},
\begin{equation}
P_{j}^{k} (\eta) = (1 - \eta^2)^{\frac{k}{2}} \frac{d^k P_{j}(\eta)}{d\eta^k},
\label{pjk}
\end{equation}
are the solutions of the differential equation
\begin{equation}
\frac{d }{d\eta} \left[(1 - \eta^2) \frac{dP_{j}^{k}}{d\eta} \right] =
\left[\frac{k^2}{1 - \eta^2} - j(j+1)\right] P_{j}^{k} (\eta), \label{dpjk}
\end{equation}
where $j$ and $k$ are integers, with $j \geq k$. On the other hand, due to the 
discontinuous character of $\eta$, $\Phi_h (\xi, \eta)$ will be continuous 
everywhere only if we take $(j - k)$ as an even number in order that $P_{j}^{k} (\eta)$ be an even function of $\eta$.

With the previous expressions, and using the Laplace operator in oblate spheroidal 
coordinates (\ref{lapop}) in the Poisson equation 
(\ref{posec}), we obtain for the mass density of the halo the expression
\begin{equation}\label{rhosum}
\varrho (\xi,\eta) = \frac{1}{4 \pi G} \sum_{j=0}^{m} \sum_{k=0}^{j}
B_{jk} \varrho_j^k (\xi,\eta), 
\end{equation}
where
\begin{equation}
\varrho_j^k (\xi,\eta) = \frac{k^2 q_{j}^{k} (\xi) P_{j}^{k} (\eta)}{a^2 (1 + \xi^2) (1 - \eta^2)}. \label{rhokj}
\end{equation}
Now, from (\ref{pjk}) and (\ref{rhokj}) it is easy to see that at the $z$ axis, when $\eta = \pm 1$, the function $\varrho_j^k (\xi,\eta)$ diverges for $k = 1$ and vanishes for $k > 2$. Accordingly, in order to have a well behaved mass density for the halo, we only consider in expression (\ref{pothgen}) the terms with $k = 0$ and $k = 2$ and so, in order to grant the continuity of the potential, we must take $j$ as an even number. Furthermore, in order to have a nonzero mass density for the halo, we must consider models with $m \geq 2$. Finally, as $P_{j}^{k} (\eta)$ is finite at the interval $- 1 \leq \eta \leq 1$ and
$q_{j}^{k} (\xi)$ goes to zero when $\xi \to \infty$, $\varrho (\xi,\eta)$
properly vanishes at infinity. 

A simple possibility for the halo potential in agreement with the above considerations is given by taking (\ref{pothgen}) with $m = 4$, \begin{align}
\Phi_h (\xi,\eta) =& \ B_{00} \ q_{0}^{0} (\xi) P_{0}^{0} (\eta)  +
B_{20} \ q_{2}^{0} (\xi) P_{2}^{0} (\eta) \nonumber \\
&\  + B_{22} \ q_{2}^{2} (\xi) P_{2}^{2} (\eta)  +
B_{40} \ q_{4}^{0} (\xi) P_{4}^{0} (\eta) \nonumber \\
&\ +
B_{42} \ q_{4}^{2} (\xi) P_{4}^{2} (\eta), \label{potenh}
\end{align}
in such a way that at least two terms in (\ref{rhosum}) contribute to the mass halo density. Then,  after using the explicit expressions for $q_{j}^{k} (\xi)$ and $P_{j}^{k}
(\eta)$, the halo density can be written as
\begin{align}
\varrho (\xi, \eta)= \frac{3}{\pi G a^2}& \left\{B_{22}\left[3\acot\xi-\frac{\xi(5+3\xi^{2})}{(1+\xi^{2})^{2}}\right] +\right. \nonumber \\
&\ 
\left.5 B_{42}(7\eta^{2}-1) \left[\frac{15}{4}(1+7\xi^{2})\acot\xi\right.\right. \nonumber \\
&\ -
\left.\left. \frac{\xi(81+190\xi^{2}+105\xi^{4})}{4(1+\xi^{2})^{2}}\right] \right\}  \label{denh}
\end{align}
which is maximum at the disc surface, when $\xi = 0$, and then fastly
decreases being constant at the oblate spheroids defined by $\xi = \rm cte$.

By
integrating over all the space, we obtain the expression
\begin{equation}
B_{22} + 6 B_{42}= \frac{\mathcal{M}_h G}{16 a}, \label{bmass}
\end{equation}
where ${\mathcal M}_h$ is the total mass of the halo. Furthermore, from the
condition (\ref{cond1a}), we obtain the relations
\begin{subequations}\label{b33}
\begin{align}
B_{22} &= - \frac{5 B_{00}}{96} + \frac{B_{20}}{48} , \label{b22} \\
B_{40} &= - \frac{3 B_{00}}{8} - \frac{3 B_{20}}{4}, \label{b40} \\
B_{42} &= - \frac{B_{00}}{576} - \frac{B_{20}}{288}. \label{b42} \\ \nonumber
\end{align}
\end{subequations}
Finally, solving the system of equations (\ref{bmass}) and (\ref{b33}), we obtain
\begin{subequations}
\begin{align}
B_{00} &= - \frac{\mathcal{M}_h G}{a}, \label{b00} \\
B_{20} &=  \frac{\mathcal{M}_h G}{2 a} - 288 B_{42} , \label{b20} \\
B_{22} &= \frac{\mathcal{M}_h G}{16a} -6 B_{42}, \label{b22} \\
B_{40} &= - 216 B_{42}, \label{b40} \\ \nonumber
\end{align}
\end{subequations}
and so all the constants in (\ref{potenh}) are expressed in terms of the halo
mass ${\mathcal M}_h$ and the coefficient $B_{42}$.

On the other hand, if we restrict to particles moving in the thin disc, the circular 
velocity is written in terms of the spheroidal coordinates as
\begin{equation}
v_c^2 = \left. \frac{(\eta^2 - 1)}{\eta} \frac{\partial \Phi}{\partial \eta}
\right|_{\xi=0},
\end{equation}
which, by using (\ref{poten}), (\ref{potend}), (\ref{potenh}) and the
properties of the Legendre functions, reduces to
\begin{equation}
v_c^2 ({\widetilde R}) = \frac{{\widetilde R}^2}{\eta} \sum_{l=1}^{m}
{\widetilde C}_{2l} P'_{2l} (\eta), \label{vcsph}
\end{equation}
where
\begin{subequations}
\begin{align}
{\widetilde C}_2 & = q_2 (0) \left[ C_2 + 66 B_{42} + \frac{\mathcal{M}_h G}{4 a} \right],\label{c2tilde} \\ 
{\widetilde C}_4 & = q_4 (0) \left[ C_4 + 24 B_{42}\right],\label{c4tilde} \\ \nonumber
\end{align}
\end{subequations}
and
\begin{equation}
{\widetilde C}_{2l} = q_{2l} (0) C_{2l}, \label{c2ltil}
\end{equation}
for $l \geq 3$. Then, by using (\ref{constant0}), (\ref{c2tilde}), (\ref{c4tilde}) and
(\ref{c2ltil}), it is easy to establish that
\begin{equation}
\frac{{\mathcal M}_d G}{a} + \frac{{\mathcal M}_h G}{4 a} = \sum_{l=1}^{m}
\frac{(-1)^{l+1}{\widetilde C}_{2l}}{q_{2l}(0)} - 42 B_{42}, \label{linear}
\end{equation}
and thus the model implies a linear relationship between ${\mathcal M}_d$ and
${\mathcal M}_h$, where the independent term is determined by the constants
${\widetilde C}_{2l}$, with $l \geq 1$, and the coefficient $B_{42}$. Now, it is clear that the above
relationship makes sense only if the right hand side it is positive, which
should be checked for every set of constants ${\widetilde C}_{2l}$ corresponding
to any particular model. The coefficient $B_{42}$ must be chosen in such a way that the model represent galaxies with a surface density mass and vertical frequency with a physically acceptable behavior.

\section*{Obtaining Particular Models}\label{partmodel}

In order to obtain particular models, we must specify the constants
${\widetilde C}_{2l}$ of the general model. So,  we will adjust these constants
in such a way that the circular velocity $v_{c}^{2}({\widetilde R})$ fits with the
data of the rotation curve of some particular galaxy. As expression
(\ref{vcsph}) for the circular velocity only involves derivatives of the
Legendre polynomials of even order, it can be written as the rotation law \citep{GPR}
\begin{equation}
v_{c}^{2}({\widetilde R}) = \sum_{l=1}^{m} A_{2l} {\widetilde R}^{2l},
\label{velr2n}
\end{equation}
where the $A_{2l}$ constants are related with the previous constants
${\widetilde C}_{2l}$, for $l \neq 0$, through the relation
\begin{equation}
{\widetilde C}_{2l} = \frac{4l + 1}{4l(2l + 1)}\sum_{k = 1}^{m} A_{2k} I_{kl},
\label{relation}
\end{equation}
where
\begin{equation}
I_{kl} = \int_{-1}^{1} \eta(1 - \eta^{2})^{k} P'_{2l}(\eta) d\eta,
\end{equation}
which is obtained by equaling expressions (\ref{vcsph}) and (\ref{velr2n})
and by using the orthogonality properties of the associated Legendre functions
\citep{AW}.

Then, if the constants $A_{2l}$ are determined by a fitting of the observational 
data of the corresponding rotation curve, the corresponding values of the 
coefficients ${\widetilde C}_{2l}$ can be determined by means of relation 
(\ref{relation}), obtaining then a particular case  of (\ref{linear}) 
corresponding to a specific galaxy model, which can be written in terms of the 
constants $A_{2l}$ as
\begin{equation}
\frac{{\mathcal M}_d G}{a} + \frac{{\mathcal M}_h G}{4 a} = \sum_{k,l=1}^{m} 
\frac{(-1)^{l+1} (4 l + 1) A_{2k} I_{kl}}{4 l (2 l + 1) q_{2l}(0)}-42 B_{42}. \label{lineara2n}
\end{equation}
However, this relation does not determine completely the values of
${\mathcal M}_d$ and ${\mathcal M}_h$, but only gives a linear relationship between 
them. So, in order to restrict the allowed values of these masses, it is needed 
to analyse the behaviour of some other quantities characterizing the kinematics 
of the model. These features are the epicycle or radial frequency, $\kappa^{2} (R)$, 
and the vertical frequency, $\nu^{2} (R)$, which describe the stability against 
radial and vertical perturbations of particles in quasi-circular orbits \citep{BT}. These  frequencies, which must be positive in order to have stable circular orbits, are defined as
\begin{align}
&\kappa^{2} (R) = \left. \frac{\partial^{2} \Phi_{\rm eff}}{\partial R^{2}}
\right|_{z=0}, \label{epicycle} \\
&	\nonumber	\\
&\nu^{2} (R) = \left. \frac{\partial^{2} \Phi_{\rm eff}}{\partial z^{2}}
\right|_{z=0}, \label{vertical}
\end{align}
where
\begin{equation}
\Phi_{\rm eff} = \Phi (R,z) + \frac{\ell^{2}}{2R^{2}}, \label{phieff}
\end{equation}
is the effective potential and $\ell = R v_c$ is the specific axial angular
momentum. Then, by using expression (\ref{vc2}) for the circular velocity,
we can write the above expressions as
\begin{align}
&\kappa^2 (R) =  \frac{1}{R} \frac{dv_c^2}{dR} + \frac{2 v_c^2}{R^2},
\label{k2} \\
&	\nonumber	\\
&\nu^2 (R) = \left. \nabla^2 \Phi \right|_{z=0} - \frac{1}{R} \frac{dv_c^2}{dR},
\label{lap0}
\end{align}
where we also used the expression for the Laplace operator in cylindrical
coordinates.

Now, by using (\ref{velr2n}), the epicycle frequency can be cast as
\begin{equation}
{\widetilde \kappa}^2 ({\widetilde R}) = \sum_{l=1}^{n} 2 (l + 1)
A_{2l} {\widetilde R}^{2l - 2}, \label{kapr2n}
\end{equation}
where ${\widetilde \kappa} = a\kappa$. It is easy to notice that the above expression is completely determined by the set of constants $A_{2l}$, which are fixed by the 
numerical fit of the rotation curve data, such that it is not possible to find a relation between the disc and halo masses that can be adjusted by requiring radial stability. On the other hand, by using the Poisson equation (\ref{posec}), the expression
(\ref{denh}) for the halo density and the expression (\ref{velr2n}) for the 
circular velocity, we find that the vertical frequency can be written as
\begin{equation}
{\widetilde \nu}^2  ({\widetilde R}) = f_{\nu}({\mathcal{M}_{h}}, B_{42},{\widetilde R})-
f_1 ({\widetilde R}), \label{nur2n}
\end{equation}
where ${\widetilde \nu} = a \nu$,
\begin{equation}
f_{\nu}({\mathcal{M}_{h}}, B_{42},{\widetilde R})= \frac{9 \pi G {\mathcal M}_h}{8 a}+567 \pi B_{42}-\frac{1575\pi}{2}B_{42}{\widetilde R}^{2}, \label{fnu}
\end{equation}
and
\begin{equation}
f_1 ({\widetilde R}) = \sum_{l=1}^{m} 2 l A_{2l} {\widetilde R}^{2l - 2}.
\label{f1}
\end{equation}
Thus, as $\widetilde \nu^{2}$ must be positive everywhere at the interval $0\leq \widetilde R \leq 1$ in order to have vertically stable models, it must satisfy that
\begin{equation}
 f_{\nu}({\mathcal{M}_{h}}, B_{42},{\widetilde R}) \geq f_{1}({\widetilde R}),\label{intervalo1}
\end{equation}
which give us a range for $\mathcal{M}_{h}$ y $B_{42}$ such that $\widetilde \nu^{2} \geq0$.

Now, we also need to consider the behavior of the surface mass density, which by using the condition (\ref{finitden}) and replacing (\ref{c2tilde}) and
(\ref{c4tilde}) in (\ref{density}), can be written as
\begin{equation}
\Sigma(\widetilde{R}) = \frac{\sqrt{1 - \widetilde{R}^{2}}}{2 \pi a G} \left\{ f_{2}(\widetilde{R})-f_{d}(\mathcal{M}_{h}, B_{42}, \widetilde{R}) \right\},
\end{equation}
where
\begin{equation}
f_{2}(\widetilde{R}) = \sum_{l=1}^{m} \frac{\widetilde{C}_{2l}(2l+1)}{\sqrt{1-\widetilde{R}^{2}}} \frac{q_{2l+1}(0)}{q_{2l}(0)}\left[\frac{P_{2l}
(\eta)-P_{2l}(0)}{\eta} \right],\label{f2}
\end{equation}
and
\begin{equation}
f_{d}(\mathcal{M}_{h}, B_{42}, \widetilde{R}) = \frac{3\mathcal{M}_{h} G}{4a} + 238 B_{42} - 280 B_{42}\widetilde{R}^{2}. \label{fd}
\end{equation}
So, in order for the surface mass density to be positive in the interval $0\leq \widetilde R \leq 1$, it must be met that
\begin{equation}
 f_{2}(\widetilde{R})\geq f_{d}(\mathcal{M}_{h}, B_{42}, \widetilde{R}).\label{intervalo2}
\end{equation}
The relation (\ref{intervalo2}) give us another range of values, not  necessarily equal to the relation (\ref{intervalo1}),
for which we obtain a  surface density mass with a acceptable behavior. Then, in order that the model make sense, we must verify that it meets the relation
\begin{equation}
  f_{\nu}({\mathcal{M}_{h}}, B_{42},{\widetilde R}) \geq f_{1}({\widetilde R})\cap  f_{2}(\widetilde{R})\geq f_{d}(\mathcal{M}_{h}, B_{42}, \widetilde{R}),
\label{condifinal}
\end{equation}
which should be checked for every set of constants corresponding to any particular model.

\section*{Adjusting Data to Models}\label{fitting}

\begin{figure*}[b]
\begin{center}
\begin{tabular}{cc}
\includegraphics[width=2.75in,height=2.75in]{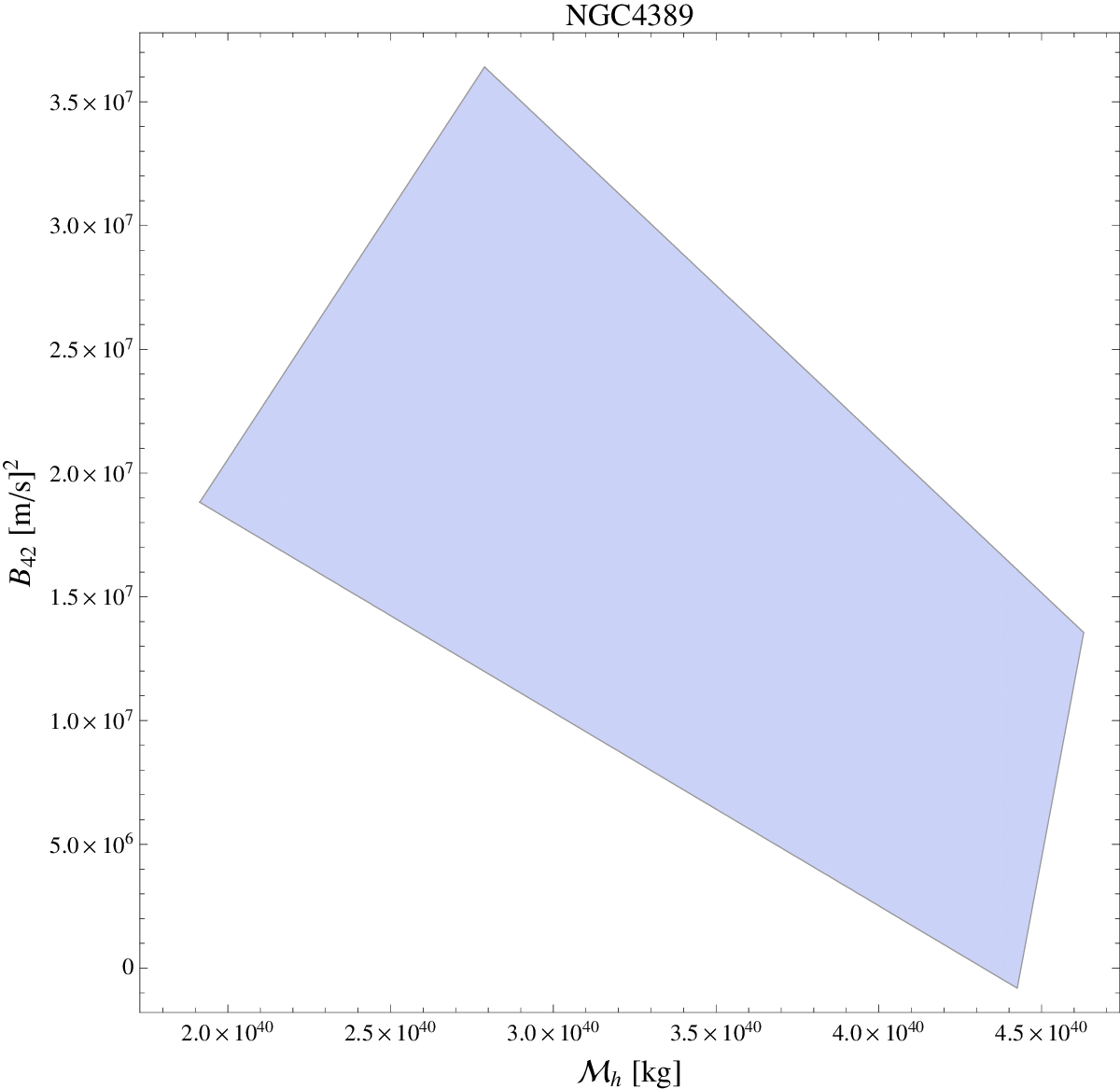} \ \ & \ \
\includegraphics[width=2.75in,height=2.75in]{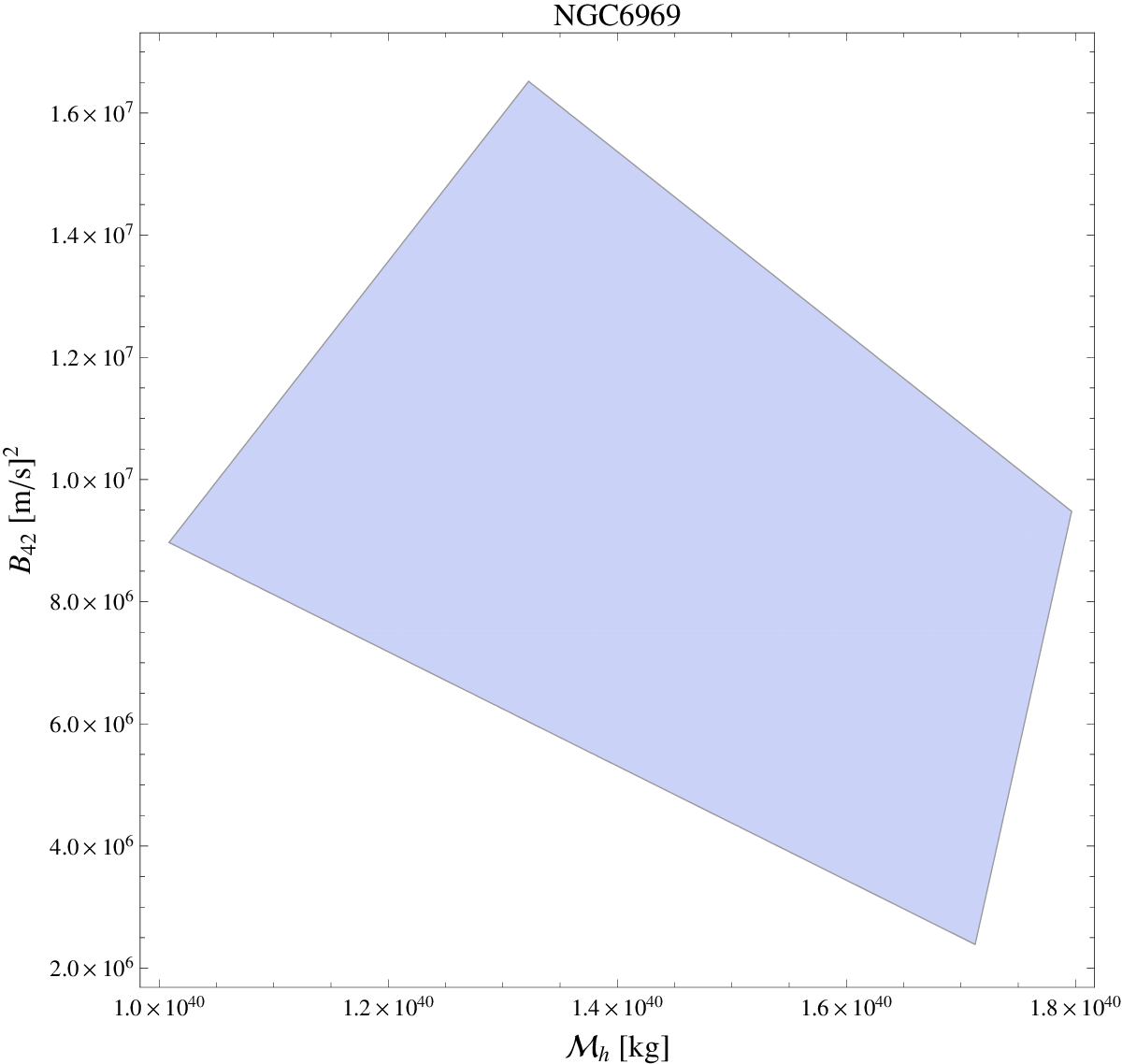} \\
\hspace{1cm}  (a) & \hspace{1cm}(b) \\
\end{tabular}
\end{center}
\caption{Solution interval of the condition (\ref{condifinal}). In (a) we show the interval of parameters $\mathcal{M}_h$ and $B_{42}$ 
for the galaxy NGC4389. In (b) we show the interval of parameters $\mathcal{M}_h$ and $B_{42}$ for the galaxy UGC6969.}
\label{interval}
\end{figure*}

In order to illustrate the above model to the real observed data, we have taken a sample of spiral galaxies of the Ursa Major cluster. We pick the corresponding data out
from Table 4 of the paper by \cite{VS}, which presents the results of 
an extensive 21 cm-line synthesis imaging survey of 41 galaxies in the nearby of the
Ursa Major cluster using the Westerbork Synthesis Radio Telescope. 
The mean distance between this telescope and the cluster is 18.6 Mpc. At this distance, 1 arcmin corresponds to 5.4 kpc. 

For each rotation curve data, we take as the value of $a$, the value given by the last tabulated 
radius, i.e. we are assuming that the radius of each galaxy is
defined by its corresponding last observed value. Although this assumption about the 
galactic radius do not agrees with the accepted standard about the edge of the 
stellar disc \citep{BM}, we will make it since we are 
assuming that all the stars moving in circular orbits at the galactic plane are 
inside the disc and that there are no stars moving outside the disc.
Thereafter we take the radii normalized in units of $a$
to fit the rotation curve of every galaxy by mean of the model (\ref{velr2n}).

The fits are made through a non-linear least squares fitting using 
the Levenberg-Marquardt algorithm, implemented internally by ROOT version 5.28 \citep{ROOT},  which minimizes the weighted sum of squares of deviations between the fit and the data. We assigned weights to the data points 
inversely proportional to the square of their errors. These errors corresponding to $2v \Delta v$ being $\Delta v$
the galaxy velocity measurement error. For each galaxy, initially we look for all the possible fits 
starting at $m = 1$ up to $m = N - 1$, with N the number of measured data pairs ($R$, $v^2$), 
hence we find a value for $m$ such that we get the minimum reduced chi square $\chi ^{2}_{r}$ (the best fit).
Now we can discard the galaxies that do not pass the reduced chi squared test with a confidence level of 95\% \citep{DATARED}. 

The $\widetilde{C}_{2l}$ constants are calculated by using the relations 
(\ref{constant0}), (\ref{c2ltil}) and (\ref{relation}). Therefore, by using  this set of constants in (\ref{f1}) and (\ref{f2}) we find for each galaxy 
the functions $f_1$ and $f_2$. 
Finally, through a routine made in Mathematica 8.0., we check for each galaxy of the sample  the validity of the condition (\ref{condifinal}). 
However, when we check the consistency of the adjust, we found that only the 
fit of the data for the galaxies NGC4389 and UGC6969 it agrees with these conditions, whereas that for all the other galaxies we found that the solution 
interval for $\mathcal{M}_h$ and $B_{42}$, given by (\ref{condifinal}), is empty.

\begin{table}[H]
\centering
\caption{Constants $A_{2l}$ in units of $10^{6} \ m^2 s^{-2}$}\label{parametros}
\begin{tabular}{ccc}
\hline
\hline
\qquad \qquad &    {NGC4389}     &  {UGC6969}  \\ 
\hline
\quad $A_{2}$ \quad & 30087.0 $\pm$ 2489.3&16387.4 $\pm$ 3322.3\\
\quad $A_{4}$ \quad & -57552.0 $\pm$ 16144.7&-46813.5 $\pm$ 22369.7\\
\quad $A_{6}$ \quad & 67317.2 $\pm$ 30484.7&71401.6 $\pm$ 43160.4\\
\quad $A_{8}$ \quad & -27760.5 $\pm$ 16936.1&-34747.1 $\pm$ 24192.1\\
\hline
\end{tabular}
\end{table}

In Table \ref{parametros} we present the values of the constants $A_{2l}$, in units of $10^{6} m^2 s^{-2}$, obtained by the numerical adjust with the rotation curve data for galaxies NGC4389 and UGC6969.
With this values for the constants, we obtain, from (\ref{lineara2n}), for the galaxy NGC4389 the relationship
\begin{equation}
\mathcal{M}_{h}+4\mathcal{M}_{d}=5.72442 \times 10^{40} - 4.2641 \times 10^{32}B_{42}, \label{masadisco4389}
\end{equation}
and for the galaxy UGC6969 the relationship
\begin{equation}
\mathcal{M}_{h}+4\mathcal{M}_{d}=2.42036 \times 10^{40} - 3.56507 \times 10^{32} B_{42},  \label{masadisco6969}
\end{equation}
where all the quantities are in $kg$.

In Fig. \ref{interval}. we present the region that represent the solution interval of the condition (\ref{condifinal}) for the galaxies  NGC4389 and UGC6969.  This region represent the values that the halo mass and the coefficient   
$B42$  can take, in order to obtain galaxy models with a vertical frequency always positive and with a surface density that has a maximum value at the disc 
centre and then decreases as $\widetilde R$ increases, vanishing at the disc edge. 

In Table \ref{masas} we present, based on the values obtained by the  condition (\ref{condifinal}) and  plotted in the Fig. \ref{interval}, the minimum and maximum values for the halo mass of each galaxy and the disc mass calculated from the relations (\ref{masadisco4389}) and (\ref{masadisco6969}), in units of $10^{10} \mathcal{M}_{\odot}$, whereas in Table \ref{coefic} we present the respectives values of the coefficient $B_{42}$, in units of $10^{6} \ m^2 s^{-2}$.

In Fig. \ref{veloc}., we show the adjusted rotation curve for these two galaxies. The points with error bars are the observations as reported in 
\cite{VS}, while the solid line are the circular velocity determined from (\ref{velr2n}) and the $A_{2l}$ parameters given by the best 
fit. As we can see, for the two galaxies we get a fairly accurate numerical adjustment with the  observational rotation curve.

\begin{table}[H]
 \centering
\caption{$\mathcal{M}_h$ and $\mathcal{M}_d$ in units of $10^{10} \mathcal{M}_{\odot}$.} \label{masas}
\begin{tabular}{ccccc}
\hline 
\hline 
&\multicolumn{2}{c}{\centering NGC4389} & \multicolumn{2}{c}{\centering NGC6969}\\
& \quad min \quad & \quad max \quad & \quad min \quad & \quad max \quad \\
\hline
\qquad $\mathcal{M}_{h}$ \qquad & \quad 0.962 \quad & \quad 2.327 \quad & \quad 0.507 \quad & \quad 0.903  \quad \\
\qquad $\mathcal{M}_{d}$ \qquad & \quad 0.065 \quad & \quad 0.378 \quad & \quad 0.036 \quad & \quad 0.137  \quad \\
\hline 
\end{tabular}
\end{table}

\begin{table}[H]
\centering
\caption{Coefficient $B_{42}$ in units of $10^{6} \ m^2 s^{-2}$.} \label{coefic}
\begin{tabular}{ccccc}
\hline 
\hline 
&\multicolumn{2}{c}{\centering NGC4389} & \multicolumn{2}{c}{\centering NGC6969}\\
& \quad min \quad & \quad max \quad & \quad min \quad & \quad max \quad \\
\hline
\qquad $B_{42}$ \qquad & \quad 18.82 \quad & \quad 13.56 \quad & \quad 8.97 \quad & \quad 9.48  \quad \\
\hline
\end{tabular}
\end{table}

\begin{figure*}
\begin{center}
\begin{tabular}{cc}
\includegraphics[width=2.75in,height=2.15in]{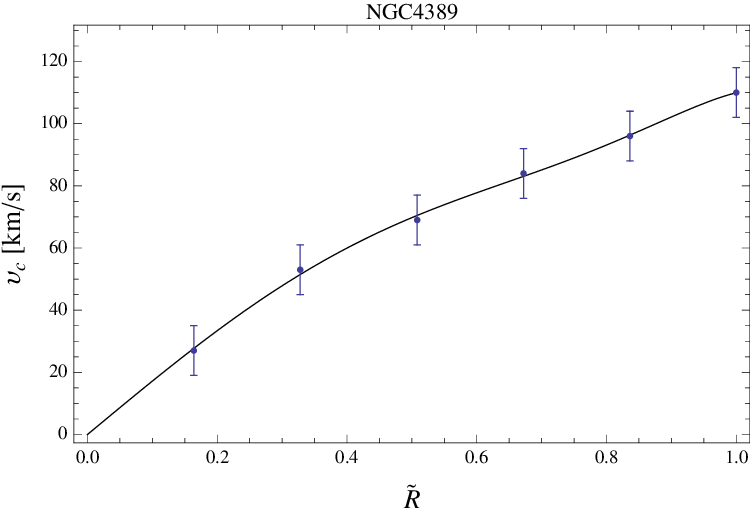} \quad & \quad 
\includegraphics[width=2.75in,height=2.15in]{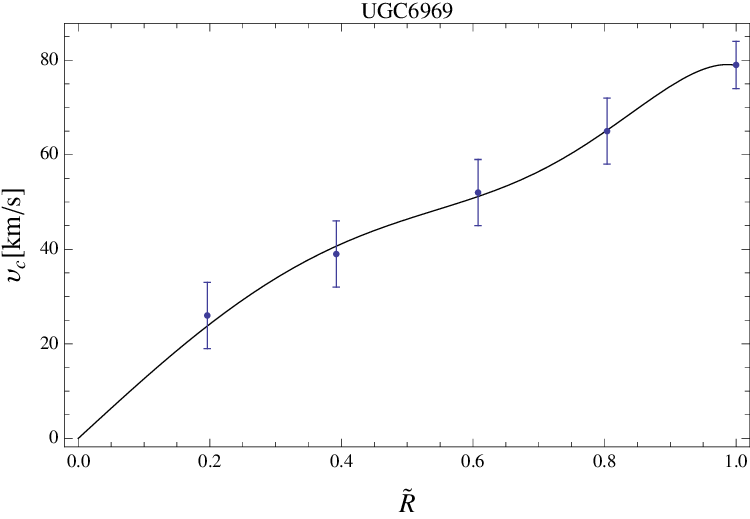} \\
\end{tabular}
\end{center}
\caption{Circular velocity $v_{c}$, as a function of 
the dimensionless radial coordinate ${\widetilde R}$, for the galaxies 
NGC4389 and UGC6969. Error bars represent the observed data by \cite{VS}, while the solid line are the circular velocity determined from  (\ref{velr2n}), and the $A_{2l}$ parameters given by the best fit.}\label{veloc}
\end{figure*}

\begin{figure*}
\begin{center}
\begin{tabular}{cc}
\includegraphics[width=2.85in,height=2.15in]{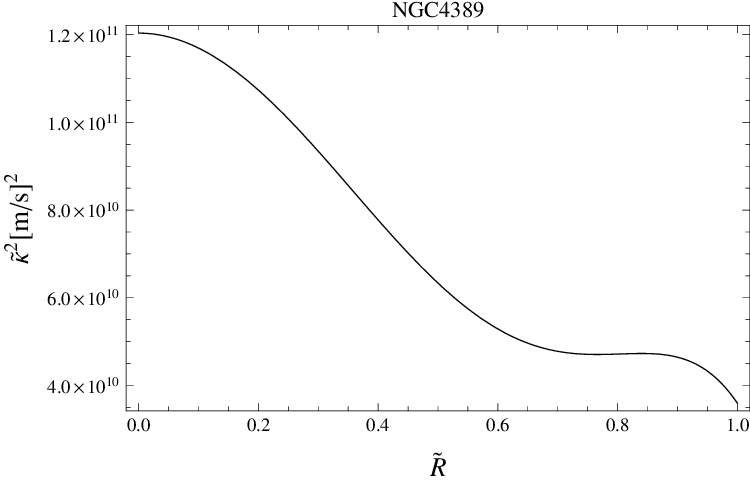} \quad & \quad 
\includegraphics[width=2.85in,height=2.15in]{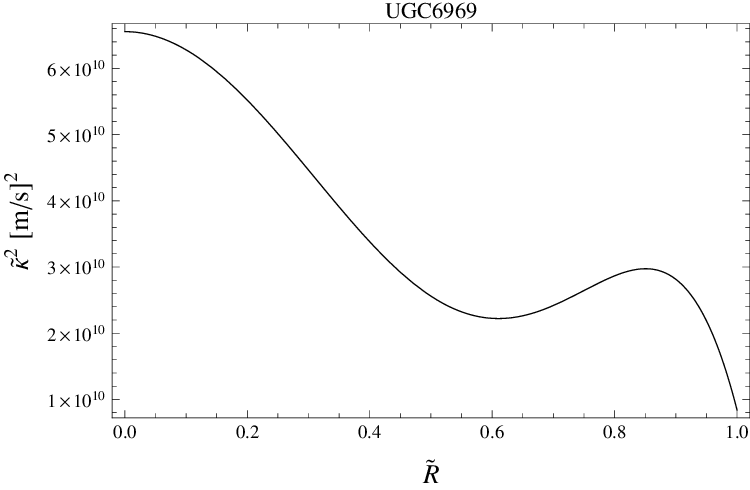} \\
\end{tabular}
\end{center}
\caption{Epicycle frequency ${\widetilde \kappa}^2 \times 10^{-3}$
in $({\rm km/s})^{2}$, as a function of the dimensionless radial coordinate
${\widetilde R} $, for the galaxies NGC4389 and UGC6969.}\label{epici}
\end{figure*}

\begin{figure*}
\begin{center}
\begin{tabular}{cc}
\includegraphics[width=2.8in,height=2.15in]{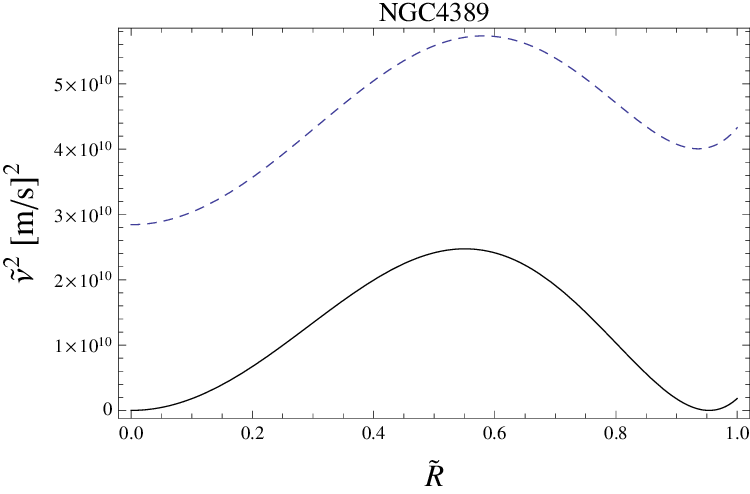} \quad & \quad 
\includegraphics[width=2.8in,height=2.15in]{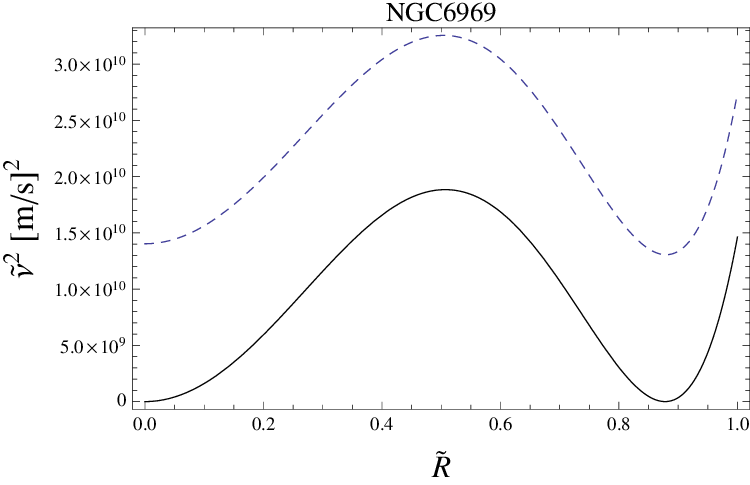} \\
\end{tabular}
\end{center}
\caption{Vertical frequency ${\widetilde \nu}^2 \times 10^{-3}$
in $({\rm km/s})^{2}$, as a function of the dimensionless radial coordinate
${\widetilde R}$, for the galaxies NGC4389 and UGC6969. The solid line represents the vertical frequency by taking
the minimum value for the halo mass, whereas the dashed line represents the vertical frequency by using the maximum value for the halo mass. }\label{verti}
\end{figure*}

In Fig. \ref{epici}. we show the 
epicycle frequency for the two galaxies. It is easy to see that this quantity is always positive, which means that the galaxies are stable against radial
perturbations.  In Fig. \ref{verti}. we present the vertical frequency for the two models. For the two galaxies, the solid line represent the vertical frequency by taking
the minimum value for the halo mass, whereas the dashed line represent the vertical frequency by using the maximum value for the halo mass. 
As can be notice in the figure, for the two galaxies the vertical frequency is positive over the entire range of $\widetilde R$, so the models are stable against 
vertical perturbations. It is easy to verify that for any other value of the halo mass and the corresponding parameter $B_{42}$, as determined from Fig. \ref{interval}., the vertical frequency remains positive in all range $\widetilde R$.

In Fig. \ref{densi}. we present the corresponding plots of the surface mass density for the two galaxies. As in the previous case, for both galaxies
the solid line represents the behavior of the surface mass density by taking the minimum value for the halo mass, while the dashed line is the surface mass
density for the maximum value of the halo mass. The behavior of this quantity is similar for both galaxies, i.e. the surface mass has a maximum value at the disc 
centre and then decreases as $\widetilde R$ increases, vanishing at the disc edge.

\begin{figure*}
\begin{center}
\begin{tabular}{cc}
\includegraphics[width=2.8in,height=2.15in]{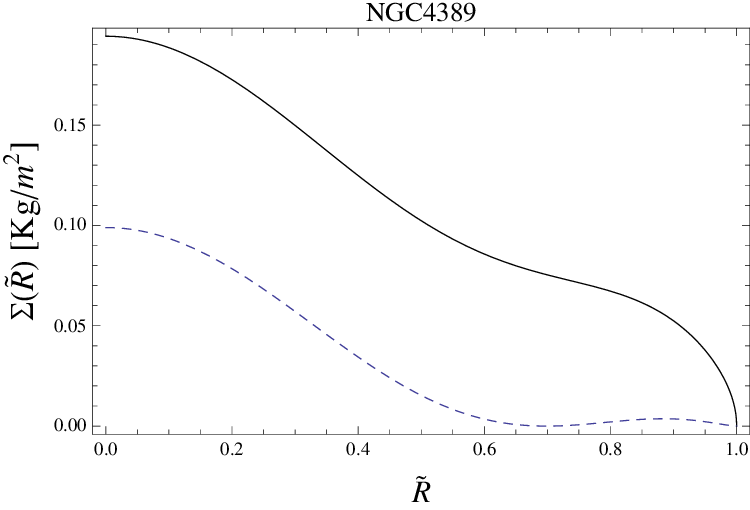} \quad & \quad 
\includegraphics[width=2.8in,height=2.15in]{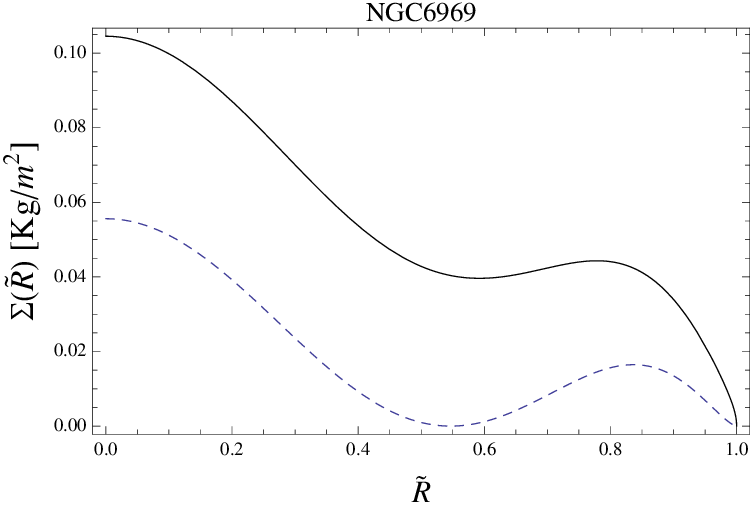} \\
\end{tabular}
\end{center}
\caption{Surface mass density ${\Sigma} \times 10^{-2}$ in
$({\rm kg/m^{2}})$, as a function of the dimensionless radial coordinate
${\widetilde R}$, for the galaxies NGC4389 and UGC6969. The solid line represents the surface mass density by taking
the minimum value for the halo mass, whereas the dashed line represents the surface mass density by using the maximum value for the halo mass. }\label{densi}
\end{figure*}

Finally,  from  (\ref{denh}), in Fig. \ref{densihalo4389}. we show the contours of the halo density distribution for the galaxy NGC4389. In plot (a), the contours are drawn using the minimum value for the halo mass, while in plot (b) we present the contours using the maximum value for the halo mass. Similarly, in Fig. \ref{densihalo6969}. we show the same quantities, but for the galaxy UGC6969. In both cases, the density profiles are positive and do not have discontinuities in all range $\widetilde R$, $z$, taking a maximum value at center and smoothly decreasing to zero when $\widetilde R \rightarrow \infty$.

\begin{figure*}
\begin{center}
\begin{tabular}{cc}
\includegraphics[width=3in,height=2.75in]{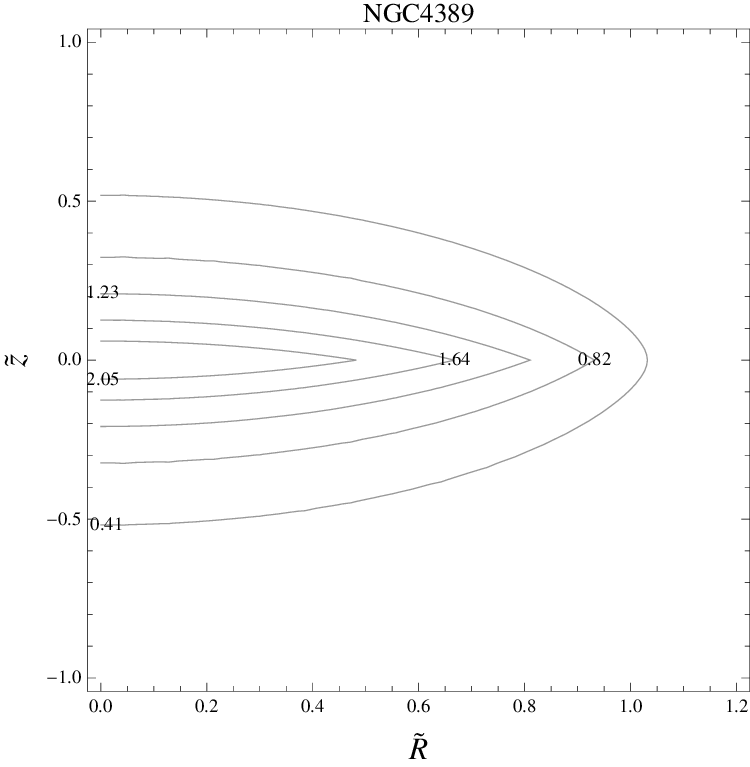} \quad & \quad 
\includegraphics[width=3in,height=2.75in]{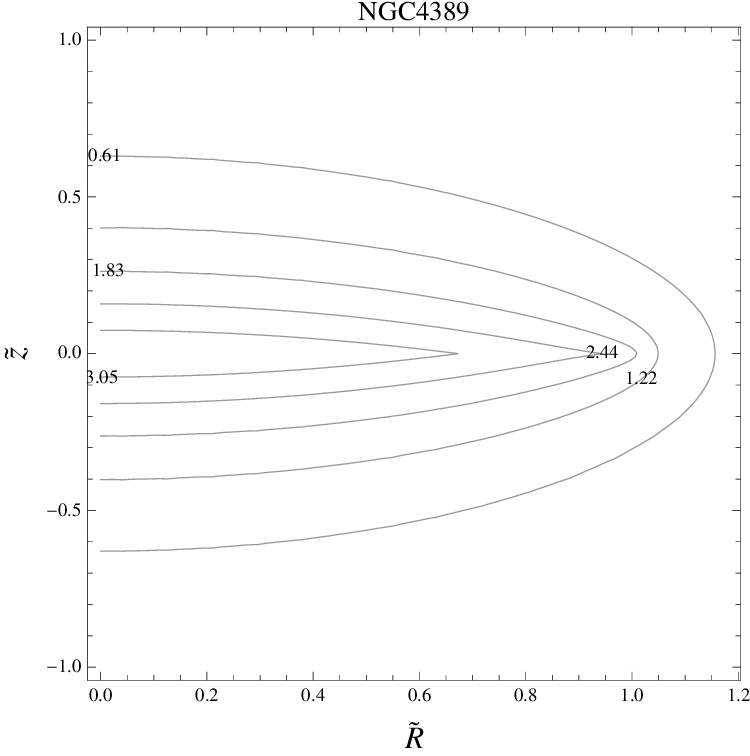} \\
\hspace{1cm}  (a) & \hspace{1cm}(b) \\
\end{tabular}
\end{center}
\caption{Contours of the halo density distribution for the galaxy NGC4389. In (a) we show the contours for the minimum value of halo mass. In (b) we show the contours using the maximum value of halo mass.}\label{densihalo4389}
\end{figure*}

\begin{figure*}
\begin{center}
\begin{tabular}{cc}
\includegraphics[width=3in,height=2.75in]{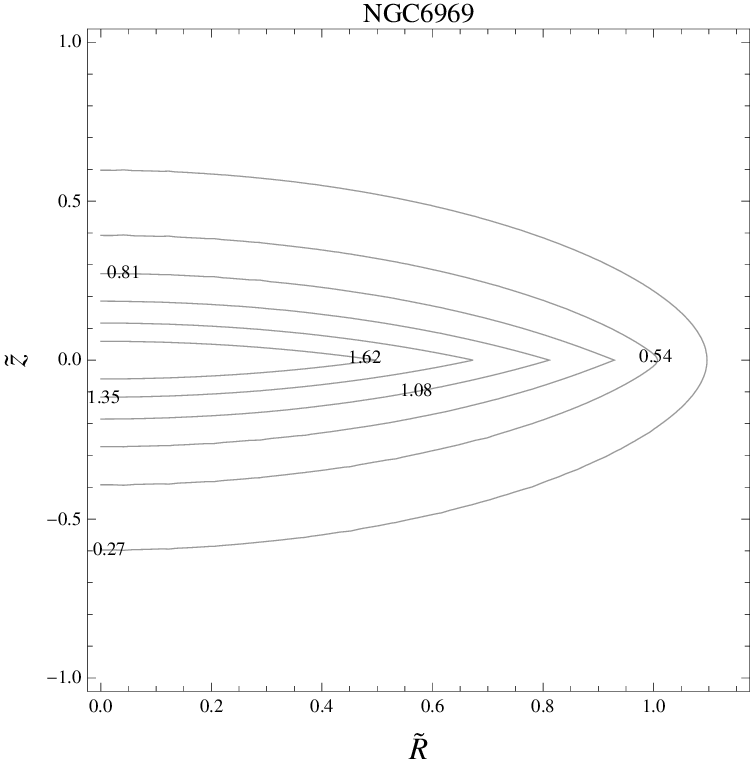} \quad & \quad 
\includegraphics[width=3in,height=2.75in]{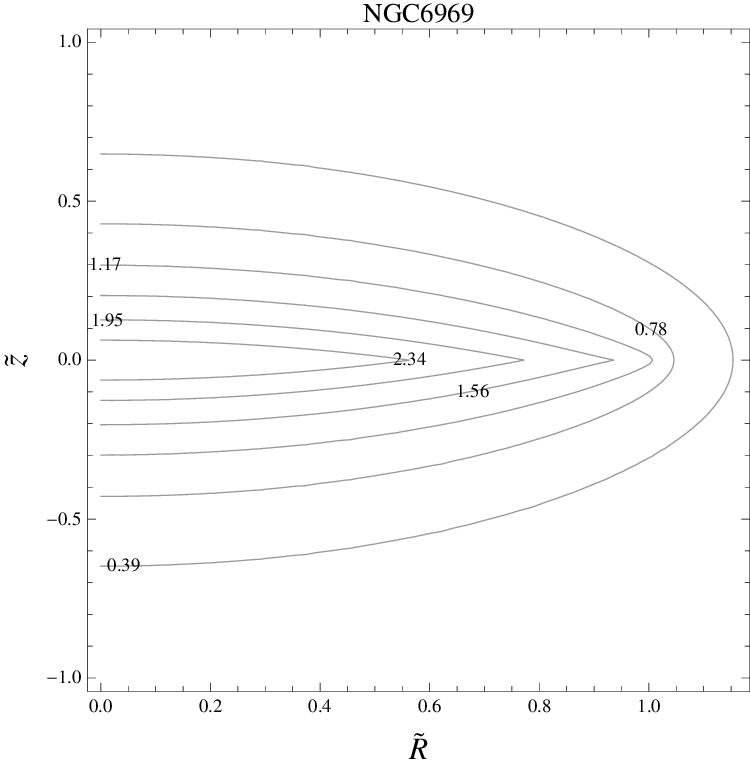} \\
\hspace{1cm}  (a) & \hspace{1cm}(b)\\
\end{tabular}
\end{center}
\caption{Contours of the halo density distribution for the galaxy UGC6969. In (a) we show the contours for the minimum value of halo mass. In (b) we show the contours using the maximum value of halo mass.}\label{densihalo6969}
\end{figure*}

\section*{Concluding Remarks}\label{discuss}

We have presented a family of analytical potentials for flat galaxies with spheroidal halos
characterised by a linear relationship between the halo mass and the disc mass. The models 
are stable against radial and vertical perturbations, and their circular velocities can be 
adjusted very accurately to the observed rotation curves of some specific galaxies. 
The here presented models are a generalisation of the models presented in 
\cite{GPR}, where only models with a thin galactic disc are considered. The 
generalisation was obtained by adding to the gravitational potential of the 
thin disc the gravitational potential corresponding to a spheroidal halo, in
such a way that we have solved the problem of vertical unstability presented by the 
previous models.

Two particular models were obtained by a numerical fit of the general expression 
(\ref{velr2n}) for the circular velocity with the observed data of the rotation 
curve of galaxies NGC4389 and UGC6969. For these two galaxies we have obtained a 
fairly accurate numerical adjustment with the rotation curve and, from the 
constants $A_{2l}$ obtained with the numerical fit, we compute the values of the 
halo mass, the disc mass and the total mass for these two galaxies in such a way 
that we obtain a very narrow interval of values for these quantities. 
Furthermore, the values of masses here obtained are in agreement with the 
expected order of magnitude, between about $10^8$ and $10^{12}$ $\mathcal{M}_{\odot}$, and with the relative order of magnitude between the 
halo mass and the disc mass, $\mathcal{M}_{d}/\mathcal{M}_{h} \approx 0.1$, \citep{ASH}. Accordingly, we believe that the values of mass obtained 
for the two studied galaxies may be taken as a very accurate estimate of the 
upper and lower bounds for the mass of the galactic disc and for the mass of the 
spheroidal halo in these two galaxies. Additionally,  the density profiles obtained  satisfy several conditions which are necessary to describe real galactic systems, i.e. they are positive and not have discontinuities in all range $\widetilde R$, $z$, taking a maximum value at center and smoothly decreasing  its value to zero when $\widetilde R \rightarrow \infty$.

However, although we tested the applicability of the present model with all the 
galaxies reported by \cite{VS}, consistent models were obtained only for the two 
galaxies NGC4389 and UGC6969, whereas for all the other galaxies were obtained 
models with values of the halo mass such that the condition (\ref{condifinal}) is not satisfied. 
Now, it can be considered that this result occurs as a consequence of
the simple halo model that we have taken here. Indeed, as we can see from expressions 
(\ref{rhokj}) and (\ref{potenh}), only one term of the gravitational potential
of the halo contributes to their density, what leaves only one free constant to 
be determined in order to fit the model to the imposed consistency conditions. 
This constant is precisely the mass of the halo, ${\mathcal M}_h$, which is 
determined by requiring the positiveness of the vertical frequency and the 
surface mass density. On the other hand, if we consider additional terms in 
expression (\ref{potenh}) for the halo potential, we will have new free 
parameters that perhaps allow to better adjust the model to properly describe the 
behavior of other galaxies besides the two considered here. 

In agreement with the above considerations, we can consider the simple set of 
models here presented as a fairly good approximation to obtaining quite realistic 
models of galaxies. In particular, we believe that the values of mass obtained 
for the two galaxies here studied may be taken as a very accurate estimate of the 
upper and lower bounds for the mass of the galactic disc and for the mass of the 
spheroidal halo in these two galaxies. Accordingly, we are now working on a more 
involved model, obtained by including additional terms in expression 
(\ref{potenh}) for the halo potential, in order to get some particular models 
that can be properly adjusted with the observed data of the rotation curve of 
some other galaxies besides the two here considered.

{\bf Acknowledgments.} The authors were supported in part by VIE-UIS, under grant number 1838, and COLCIENCIAS, Colombia, under grant number 8840. JIR wants to thank the support from Vicerrectoría Académica, Universidad Santo Tomás, Bucaramanga.

{\bf Conflict of interest.} The authors declare that they have no conflict of interest.

\bibliographystyle{chicago}

\renewcommand{\refname}

\end{multicols}
\end{small}
\end{document}